\documentstyle[prl,aps,floats,epsf]{revtex}

\begin{document}

\draft

\title{Cluster algorithms for general-$\bf S$ quantum spin systems}

\author{Synge Todo\cite{Todo} and Kiyoshi Kato\cite{Kato}}

\address{Institute for Solid State Physics, University of Tokyo, Kashiwa
277-8581, Japan}

\date{November 5, 1999}

\maketitle

\begin{abstract}
 We present a general strategy to extend quantum cluster algorithms for
 $S=\frac{1}{2}$ spin systems, such as the loop algorithm, to systems
 with arbitrary size of spins.  In general, the partition function of a
 high-$S$ spin system is represented in terms of the path integral of a
 $S=\frac{1}{2}$ model with special boundary conditions in the
 imaginary-time direction.  We introduce additional graphs to be
 assigned to the boundary part and give the labeling probability
 explicitly, which completes the algorithm together with an existing
 $S=\frac{1}{2}$ cluster algorithm.  As a demonstration of the
 algorithm, we simulate the integer-spin antiferromagnetic Heisenberg
 chains.  The magnitude of the first excitation gap is estimated as to
 be 0.41048(6), 0.08917(4), and 0.01002(3) for $S=1$, 2, and 3,
 respectively.
\end{abstract}


The world-line quantum Monte Carlo (QMC) method is one of the most
powerful tools in numerical investigations of quantum spin
systems~\cite{Suzuki93}.  One of the main advantages of the QMC method
over other numerical ones, such as exact diagonalization and density
matrix renormalization group (DMRG) method, is that it is applicable to
rather large systems in any dimensions and can estimate their physical
quantities {\em statistically exactly}.  However, the conventional
algorithm, based on local updates of spin configurations (world lines),
suffers from strong correlations between successive configurations at
low temperatures or in the vicinity of a second-order phase transition.
The diverging auto-correlation time virtually makes simulations slower
and slower, and finally it becomes practically impossible to simulate
larger systems at lower temperatures.  This drawback is called {\em
critical slowing down}.

Recently, the inventions of the {\em loop
algorithm}~\cite{EvertzLM93,WieseY94} and of its continuous-time
variant~\cite{BeardW96} have led a great improvement of the QMC
techniques for the $S=\frac{1}{2}$ XXZ model~\cite{Evertz98}.  The loop
algorithm, which is a kind of cluster algorithms, realizes updates of
the configuration by flipping non-local objects, referred to as {\em
loops}.  It has been shown that it is fully ergodic and drastically
reduces the auto-correlation time, often by orders of magnitude,
especially at low temperatures.  Furthermore, by using the
continuous-time version of the algorithm, one can completely eliminate
discretization error originating from the Suzuki-Trotter decomposition;
simulations can be performed directly in the so-called Trotter limit.

In general, it is a highly non-trivial task to construct an efficient
cluster algorithm for a given system, because symmetry or special
properties of the target system should be taken into account explicitly
in its construction.  As for the spin systems, the development of
cluster algorithms for higher-$S$ models remains as an important and
challenging problem.  A cluster algorithm for the general-$S$ XXZ model
in the discrete-time formulation has been proposed by Kawashima and
Gubernatis\cite{KawashimaG94}.  Unfortunately, their algorithm is rather
complicated (105 different graphs appear even in the $S=1$ case), and
moreover the Trotter limit is not well-defined in the algorithm.
Another kind of algorithm in the discrete-time representation has been
used in Ref.~\cite{KimGWB98}.

More recently, Harada {\it et al.} proposed a continuous-time loop
algorithm for the $S=1$ antiferromagnetic Heisenberg
model~\cite{HaradaTK98}, in which the $S=1$ system is mapped into a path
integral of an $S=\frac{1}{2}$ system with special boundary conditions
in the imaginary-time direction.  In the present letter, we generalize
their method to construct cluster algorithms for systems with arbitrary
size of spins.
For simplicity, we consider a spin-$S$ antiferromagnetic Heisenberg
model on a bipartite lattice as an example.  Generalization to other
models is straightforward as discussed later.

We consider the Hamiltonian of $N_{\text{s}}$ spins and $N_{\text{b}}$
bonds, defined as follows:
\begin{equation}
  \label{eqn:hm-1}
  {\cal H} = - \sum_{\langle i,j \rangle}^{N_{\text{b}}}
  \left( S_i^x S_j^x + S_i^y S_j^y - S_i^z S_j^z \right),
\end{equation}
where $S_i^{\alpha}$ ($\alpha=x$, $y$, $z$ and $i=1$, $\cdots$,
$N_{\text{s}}$) is the $\alpha$-component of the spin-$S$ operator at
site~$i$.

In order to construct a cluster algorithm for $S>\frac{1}{2}$, it is
crucial to represent each spin operator $S_i^\alpha$ as a sum of
$S=\frac{1}{2}$ Pauli operators $\{\sigma_{i,\mu}^\alpha\}$.  Following
Ref.~\cite{KawashimaG94}, we substitute $S_i^\alpha = \frac{1}{2}
\sum_{\mu=1}^{2S} \sigma_{i,\mu}^\alpha$ into the
Hamiltonian~(\ref{eqn:hm-1}), which yields the following Hamiltonian of
$2SN_{\text{s}}$ spins and $4S^2N_{\text{b}}$ bonds:
\begin{equation}
  \label{eqn:hm-2}
  \tilde{\cal H} = - \frac{1}{4} \sum_{\langle i,j \rangle}^{N_{\text{b}}}
  \sum_{\mu,\nu=1}^{2S}
  \left(
  \sigma_{i,\mu}^x \sigma_{j,\nu}^x + 
  \sigma_{i,\mu}^y \sigma_{j,\nu}^y -
  \sigma_{i,\mu}^z \sigma_{j,\nu}^z \right).
\end{equation}
We refer to the $S=\frac{1}{2}$ spins in Eq.~(\ref{eqn:hm-2}) as {\em
subspins} hereafter.  In terms of the Hamiltonian (\ref{eqn:hm-2})
defined on the extended phase space, the partition function of the
original Hamiltonian (\ref{eqn:hm-1}) can be expressed as
\begin{equation}
  \label{eqn:pf-1}
  Z = \mbox{Tr} ( \exp [ -\beta \tilde{\cal H} ] P ) \,.
\end{equation}
Here, we introduce a projection operator $P$, which is a direct product
of local symmetrization operators $\left\{P_i\right\}$
($i=1,\cdots,N_{\rm s}$).  Each $P_i$ acts on $2^{2S}$-dimensional
Hilbert space spanned by $\{\sigma^z_{i,\mu}\}$ ($\mu=1,\cdots,2S$), and
projects out unphysical states with ${\bf S}_i^2 < S(S+1)$.  Note that
$P$ commutes with $\tilde{\cal H}$, because $\tilde{\cal H}$ is
invariant under the exchange of subspin indices at each site by
definition.

Applying a Suzuki-Trotter decomposition for the exponential operator in
Eq.~(\ref{eqn:pf-1}) and inserting complete sets of eigenstates of
$\{\sigma_{i,\mu}^z\}$ between the exponential factors, we obtain the
following path-integral representation of the partition function:
\begin{equation}
  \label{eqn:pf-2}
  Z = \sum_{\cal C} W ({\cal C}) \, P (\partial
  {\cal C}) \ ,
\end{equation}
where ${\cal C}=\{{\cal C}_{i,\mu,\tau}\}$ ($i=1,\cdots,N_{\rm s}$,
$\mu=1,\cdots,2S$, and $0\le\tau\le\beta$) denotes a world-line
configuration of subspins and $\partial {\cal C}=\{{\cal
C}_{i,\mu,0},{\cal C}_{i,\mu,\beta}\}$ ($i=1,\cdots,N_{\rm s}$ and
$\mu=1,\cdots,2S$) denotes a configuration of subspins at the boundaries
in the imaginary-time direction.  We take the continuous-imaginary-time
representation, and ${\cal C}_{i,\mu,\tau}$ denotes the $\mu$-th
subspin direction (+1 or - 1) at $i$-th site and imaginary time $\tau$.
The weight $P(\partial {\cal C}) = \prod_i P_i(\partial {\cal C}_i)$
originates from the projection operator $P$, which can be interpreted as
{\em soft} boundary conditions in the imaginary-time direction.  Each
$(4S)$-body local boundary weight $P_i(\partial {\cal C}_i)$ takes a
value of the inverse of the number of different configurations which are
connected with $\partial{\cal C}_i$ by permutation operations, i.e.,
\begin{eqnarray}
 P_i(\partial{\cal C}_i)
 = \left\{
  \begin{array}{ll}
    \displaystyle \frac{(2S-n_i)! n_i!}{(2S)!} & 
     \mbox{if $n_{i,0} = n_{i,\beta}$ ($\equiv n_i$)} \\
    0 & \mbox{otherwise,}
  \end{array}
  \right.
\end{eqnarray}
where $n_{i,\tau}=\sum_\mu {\cal C}_{i,\mu,\tau}$.

Note that apart from the boundary conditions, the weight $W({\cal C})$
is completely equivalent to that appears in the path integral
representation of the system described by the
Hamiltonian~(\ref{eqn:hm-2}).  Therefore, for that part, we adopt the
same labeling rule as the original $S=\frac{1}{2}$ continuous-time loop
algorithm~\cite{BeardW96}, which assigns a `graph'
${\cal G}_W$ to a configuration ${\cal C}$ with labeling probability
\begin{equation}
  \label{eqn:prob-t}
  T_W ({\cal G}_W|{\cal C}) = 
  \frac{V_W ({\cal G}_W)\Delta_W({\cal C},{\cal G}_W)}{W ({\cal C})} \,.
\end{equation}
Here, we follow the general framework of cluster algorithms
presented in Ref.~\cite{KawashimaG95}.  The weight $V_W({\cal G}_W)$
is ${\cal C}$-independent and non-negative, and $\Delta_W({\cal C},{\cal
G}_W)$ is a compatibility function, which takes 0 or 1.  They satisfy
\begin{equation}
  W ({\cal C}) = \sum_{{\cal G}_W} V_W({\cal G}_W) \Delta_W({\cal
  C},{\cal G}_W)
\end{equation}
for any $\cal C$.  By this procedure the $(d+1)$-dimensional space is
decomposed into a set of loops.  Note that at this stage, some of the
loops remain opened, because we have not defined any graphs for the
boundary part.

\begin{figure}
 \begin{center} 
  \leavevmode
  \epsfxsize=0.4\textwidth\epsfbox{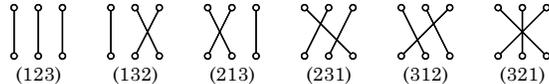}
  \vspace*{1em}
  \caption{Six graphs for the local boundary weight for $S=\frac{3}{2}$.
  The upper (lower) three circles denote subspins at $\tau=0$~($\beta$).
  In general, $(2S)!$ graphs can be labeled uniquely in terms of
  corresponding permutations of $2S$ integers as shown in the bottom
  row.}
  \label{fig:graph}
 \end{center}
\end{figure}

Next, for each local boundary weight $P_i(\partial {\cal C}_i)$, we
introduce
$(2S)!$ types of graphs $\{{\cal G}_{Pi}\}$ ($i=1,\cdots,N_{\rm s}$),
each of which consists of $2S$ edges connecting one of subspins at
$\tau=0$ to a subspin at $\tau=\beta$ one by one (Fig.~\ref{fig:graph}).
We define a compatibility function $\Delta_P(\partial {\cal C}_i,{\cal
G}_{Pi})$ as it takes 1 if every edge connects two subspins which have
an identical direction, or it takes 0 otherwise.  By using this
compatibility function, we can decompose the local boundary weight as
\begin{equation}
 P_i(\partial {\cal C}_i) = \sum_{{\cal G}_{Pi}} 
  \Delta_P(\partial {\cal C}_i,{\cal G}_{Pi})/(2S)!  \,,
\end{equation}
because the number of compatible graphs for a configuration
$\partial{\cal C}_i$ is given by $(2S-n_i)!n_i!$.  We take the labeling
probability for the local boundary weight as
\begin{equation}
 T_P({\cal G}_{Pi}|\partial{\cal C}_i) 
   =  \frac{\Delta_P(\partial {\cal C}_i,{\cal G}_{Pi})/(2S)!}{P_i(\partial {\cal C}_i)} 
   =  \frac{\Delta_P(\partial {\cal C}_i,{\cal G}_{Pi})}{(2S-n_i)!n_i!} \,,
\end{equation}
that is, a graph is assigned out of the graphs compatible with
$\partial{\cal C}_i$ with equal probability.
These boundary graphs make the remaining opened loops to be closed.  By
choosing the flipping probability to be {\it free}, that is, flipping
spins on each loop simultaneously with probability $\frac{1}{2}$, one
can show that the present stochastic process satisfies the
detailed-balance condition~\cite{KawashimaG95}.

The present algorithm includes the loop algorithm for the $S=1$
antiferromagnetic Heisenberg model by Harada {\it et
al}~\cite{HaradaTK98} as a special case.  However, as already seen, the
present strategy does not depend so much on details of the model one
considers; one can easily construct a cluster algorithm, if the mapped
$S=\frac{1}{2}$ model has a cluster algorithm, which covers a model with
general XYZ interaction or single-ion anisotropy~\cite{TodoK}, and the
transverse-field Ising model~\cite{RiegerK98}.  It can be applied also
for a system with random size of spins and even for the classical Ising
model~\cite{SwendsenW87,TodoK2000}.  The resulting general-$S$ algorithm
is ergodic, if the $S=\frac{1}{2}$ cluster algorithm lying at the base
is ergodic.  The details of the algorithm for these models and the proof
of its ergodicity will be presented elsewhere~\cite{TodoK}.  Note that
the number of graphs introduced for the local boundary weight increases
quite rapidly as $S$ increases.  However, the computational time in
selecting a graph to be assigned is merely proportional to $S$, because
the procedure is nothing but the random-permutation
generation\cite{Knuth98}.

As an application of the algorithm, we simulated the antiferromagnetic
Heisenberg chains with $S=1$, 2, and 3.  It is conjectured by
Haldane~\cite{Haldane83} that the antiferromagnetic Heisenberg chain of
integer spins has a finite excitation gap $\Delta(S)$ above its unique
ground state, and the antiferromagnetic spin correlation along the chain
decays exponentially with a finite correlation length $\xi_x(S)$.  For
$S=1$ and 2, a number of numerical studies have been accomplished (e.g.,
see \cite{WhiteH93,GolinelliJL94,SchollwoeckJ95,WangQY99}) to confirm
the validity of Haldane's conjecture.  However, estimation of the first
excitation gap of higher-spin chains has not yet been successful, since
the magnitude of $\Delta(S)$ is considered to become exponentially
small as $S$ increases~\cite{Haldane83}.

Consider a spin-$S$ chain of $L$ sites at temperature $T$ ($=1/\beta$).  We
assume $L$ is even.  The correlation function of the staggered
magnetization in the imaginary-time direction:
\begin{equation}
  \label{eqn:corr}
  C(\tau;L,\beta) = \frac{1}{L^2\beta} \sum_{i,j=1}^L \int_0^\beta \! dt \,
  \langle (-1)^{|i-j|} S_i^z(t) S_j^z(t+\tau) \rangle 
\end{equation}
is an even function of $\tau$, and satisfies $C(\tau+\beta;L,\beta) =
C(\tau;L,\beta)$.  At sufficiently low temperatures, the correlation
function is expressed well as a sum of exponential functions, that is,
\begin{equation}
  \label{eqn:sum}
  C(\tau;L,\beta) = 
  \sum_{i=0} c_i \cosh \left[ \frac{\tau-\beta/2}{\xi_{\tau,i}(L)}
  \right] \ \ \ \mbox{for $\beta\gg\xi_{\tau,0}$.}
\end{equation}
Here, we assume $\xi_{\tau,0} > \xi_{\tau,1} > \xi_{\tau,2} > \cdots$
without loss of generality.  The coefficients $\left\{c_i\right\}$ are
directly related to the dynamic structure factor at momentum $k=\pi$.
In terms of $\xi_{\tau,0}(L)$, the gap to be estimated
is given by
\begin{equation}
  \Delta = \lim_{L\rightarrow\infty} 
  \frac{1}{\xi_{\tau,0}(L)} \ .
\end{equation}

\begin{figure}
 \begin{center} 
  \leavevmode
  \epsfxsize=0.45\textwidth\epsfbox{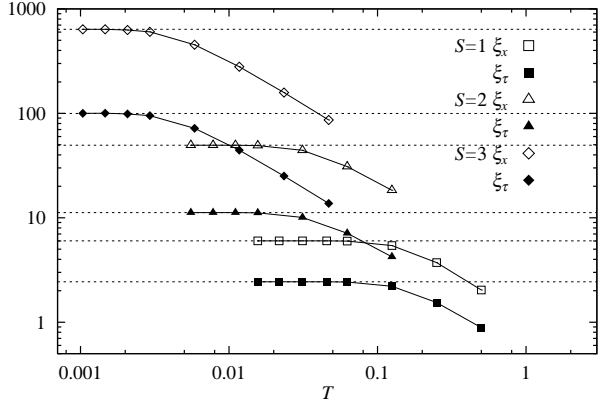}
  \caption{Temperature dependence of the spatial correlation length
  $\xi_x$ (open symbols) and the temporal correlation length $\xi_\tau$
  (solid symbols) of the $S=1$ (squares), 2 (triangles), and 3
  (diamonds) antiferromagnetic Heisenberg chains.  Statistical error of
  each data point is smaller than the line width.  The system size is
  taken as $L=2S/T$.}
  \label{fig:correlation}
 \end{center}
\end{figure}

\begin{table}[b]
  \caption{Ground-state energy density $E/L$, staggered susceptibility
  $\chi_{\rm s}$, spatial correlation length $\xi_x$, and first excited
  gap $\Delta$ of the $S=1$, 2, and 3 antiferromagnetic Heisenberg
  chains.  Note that for each $S$, the physical quantities are estimated
  by a single Monte Carlo run on the system of size $L$ at temperature $T$, which are
  presented in the second and third columns, respectively.}
  \label{tab:result}
  \begin{tabular}{crlcllll}
   $S$ & $L$\ \ & \multicolumn{1}{c}{$T$} & MCS & 
   \multicolumn{1}{c}{$E/L$} & \multicolumn{1}{c}{$\chi_{\rm s}$} & 
   \multicolumn{1}{c}{$\xi_x$} & \multicolumn{1}{c}{$\Delta$} \\
   \hline
   1 &  128 & 0.0156250 & $2\times10^7$ 
   & \ \,-1.401481(4) & \ \,\ \,\ \,\ \,18.4048(7) 
   & \ \,\ \,6.0153(3) & 0.41048(6) \\
   2 &  724 & 0.0055249 & $2\times10^6$ 
   & \ \,-4.761249(6) & \ \,\ \,1164.0(2) 
   & \ \,49.49(1) & 0.08917(4) \\
   3 & 5792 & 0.0010359 & $3\times10^4$ 
   & -10.1239(1) & 158000(310) 
   & 637(1) & 0.01002(3) \\
  \end{tabular}
\end{table}

In general, to solve Eq.~(\ref{eqn:sum}) directly is extremely
ill-posed~\cite{SilverSG90}.  However, as shown below, we can construct
a systematic series of estimators at least for $\xi_{\tau,0}(\tau)$, if
the coefficient $c_i$ in Eq.~(\ref{eqn:sum}) converges to zero rapidly
enough for large $i$, and also if the difference between
$\xi_{\tau,0}(L)$ and $\xi_{\tau,1}(L)$ remains finite even in the
thermodynamic limit.  For S=1, these two conditions are numerically
shown to be satisfied~\cite{Takahashi94}.  We expect similar situations
for higher $S$, although there is no exact argument.

For a given $L$, the well-known second-moment
estimator~\cite{CooperFP82} of the correlation length,
\begin{equation}
 \label{eqn:second-moment}
 \hat{\xi}_\tau^{(2)} = \frac{\beta}{2\pi} \sqrt{
  \frac{{\tilde{C}}(0)}{\tilde{C}(2\pi/\beta)}-1} ,
\end{equation}
converges to $\xi_{\tau,0}(L)$ in the low-temperature limit, besides
systematic corrections of $O(a_i \xi_{\tau,i}/\xi_{\tau,0})$ ($i=1, 2,
\cdots$).  Here, ${\tilde{C}}(\omega)$ is the Fourier transform of the
imaginary-time spin correlation function, i.e.,
$\tilde{C}(\omega)=\int_0^\beta C(\tau) e^{i\omega\tau} d\tau$.  In the
present algorithm, $\tilde{C}(\omega)$ can be measured directly by means
of the improved estimator\cite{BakerK95}, which reduces the variance of
the data greatly.  We also consider a {\it fourth-moment} estimator:
\begin{equation}
  \hat{\xi}_\tau^{(4)} = \frac{\beta}{4\pi} 
  \sqrt{3 \frac{{\tilde{C}}(0)-{\tilde{C}}(2\pi/\beta)}%
    {{\tilde{C}}(2\pi/\beta)-{\tilde{C}}(4\pi/\beta)}-1} \,,
  \label{eqn:fourth-moment}
\end{equation}
which has smaller corrections of $O(a_i (\xi_{\tau,i}/\xi_{\tau,0})^3)$.
Construction of higher-order estimators is possible in a straightforward
way\cite{TodoK}.

The temperature dependence of $\hat{\xi}_\tau(L,T)$ and also of the
correlation length along the chain, $\hat{\xi}_x(L,T)$, is shown in
Fig.~\ref{fig:correlation}.  In the present simulation, the system size
is taken as $L=2S/T$.  The integrated auto-correlation time of
$\tilde{C}(0)$ is of order unity and no significant sign of its growth
is observed for larger $L$ and $1/T$.  Measurement of physical
quantities is performed after discarding first $10^3$ Monte Carlo steps
(MCS) for thermalization.  One Monte Carlo step of the $S=3$ chain with
$L=5792$ and $T=0.0010359$, which is mapped to the system of 34752
subspins and 208512 bonds prior to the simulation, takes about 16
seconds on 256 nodes of Hitachi SR-2201.

As seen clearly in Fig~\ref{fig:correlation}, $\hat{\xi}_\tau(L,T)$ and
$\hat{\xi}_x(L,T)$ converge quite rapidly (probably exponentially) to a
finite value for small $T$.  We observe that for each $S$ the data which
satisfy the conditions, $1/T>6\hat{\xi}_\tau$ and $L>6\hat{\xi}_x$,
exhibit no temperature (and system-size) dependence besides statistical
fluctuations.  The difference between the second-moment estimate
(Eq.~(\ref{eqn:second-moment})) and the fourth-moment one
(Eq.~(\ref{eqn:fourth-moment})) is invisible in the vertical scale of
Fig~\ref{fig:correlation}.  The difference between these two estimates
at the lowest temperature is about 0.2\% and 0.1\% for $S=1$ and $S=2$,
respectively.  For $S=3$, both coincide within the statistical errors.
Furthermore, it is confirmed that there is no significant difference
between the fourth-moment estimate and the sixth-moment one even for
$S=1$.

Thus, by using the fourth-moment estimator, we conclude
\begin{equation}
  \Delta(S) = \left\{
  \begin{array}{ll}
    0.41048(6) & \mbox{\ for $S=1$} \\
    0.08917(4) & \mbox{\ for $S=2$} \\
    0.01002(3) & \mbox{\ for $S=3$}
  \end{array}
  \right.
\end{equation}
as the magnitude of the Haldane gap.  The results for other physical
quantities, such as the energy density and the staggered susceptibility,
are presented in Table~\ref{tab:result}.  It should be emphasized that
the present results are obtained without any extrapolation procedure;
they are simply obtained by a single Monte Carlo run on the largest
system at the lowest temperature for each $S$.

For $S=1$, the present estimate is completely consistent with
$0.41050(2)$ and $0.41049(2)$ obtained by the DMRG
calculation~\cite{WhiteH93} and by the exact
diagonalization~\cite{GolinelliJL94}, respectively.  For $S=2$, on the
other hand, the numerical uncertainty in the present estimate is much
smaller than in the previous studies (see, e.g., TABLE~I in
Ref.~\cite{WangQY99}).  Furthermore, our estimate is slightly larger
than $0.0876(13)$, which is obtained by the most recent DMRG
calculation~\cite{WangQY99}.  In the DMRG study, for some technical
reasons, open boundary conditions have been used, which is known to give
quite large systematic corrections compared to the periodic boundary
conditions.  The reason of the disagreement might be due to an
inappropriate scaling assumption in the DMRG study.  Finally, as for the
$S=3$ case, it might be practically impossible to estimate the value of
the gap by other numerical methods.  The present result is completely
new to our best knowledge.

The present authors would like to thank H. Takayama, N. Kawashima, H.~G. 
Evertz, and K. Hukushima for stimulating discussions and comments.  Most
of numerical calculations for the present work have been performed on
the CP-PACS at University of Tsukuba, Hitachi SR-2201 at Supercomputer
Center, University of Tokyo, and the RANDOM at MDCL, Institute for Solid
State Physics, University of Tokyo.  The present work is supported by
the ``Large-scale Numerical Simulation Program'' of Center for
Computational Physics, University of Tsukuba, and also by the ``Research
for the Future Program'' (JSPS-RFTF97P01103) of Japan Society for the
Promotion of Science.

\cleardoublepage

\end{document}